\documentstyle[11pt,newpasp,twoside,epsf]{article}
\def\edcomment#1{\iffalse\marginpar{\raggedright\sl#1\/}\else\relax\fi}
\reversemarginpar

\begin{document}
\title{The Extra-Solar Planet Imager (ESPI)}
  \author{P. Nisenson, G.J. Melnick, J. Geary, M. Holman, S.G. Korzennik, R.W. Noyes, C. Papaliolios, D.D. Sasselov}
\affil{Harvard-Smithsonian Center for Astrophysics, 60 Garden St., Cambridge, MA 01803}
\author{D. Fischer}
\affil{University of California, 601 Campbell Hall, Berkeley, CA 94720}
\author{D. Gezari, R.G. Lyon}
\affil{NASA GSFC, Greenbelt MD 20771}
\author{R. Gonsalves}
\affil{Tufts University}
\author{C. Hardesty}
\affil{Optical Design Services, North Reading, MA}
\author{M. Harwit}
\affil{511 H Street, SW, Washington, D.C. 20024 }
\author{M.S. Marley}
\affil{NASA Ames Research Center, Moffett Field, CA 94035}
\author{D.A. Neufeld}
\affil{Johns Hopkins University, Homewood Campus, Baltimore, MD 21218}
\author{S.T. Ridgway}
\affil{NOAO, 950 N. Cherry St, Tucson, AZ 85726}

\begin{abstract}
ESPI has been proposed for direct imaging and
spectral analysis of giant planets orbiting solar-type stars. ESPI
extends the concept suggested by Nisenson and Papaliolios (2001) for a
square aperture apodized telescope that has sufficient dynamic range
to directly detect exo-planets.  With a 1.5 M square mirror, ESPI can
deliver high dynamic range imagery as close as 0.3 arcseconds to
bright sources, permitting a sensitive search for exoplanets around
nearby stars and a study of their characteristics in reflected light.

\end{abstract}

\section{Introduction}

Since the first detection of a planet orbiting another star, the
presence of more than 100 other extra-solar planets has been inferred
from the small reflex motions that they gravitationally induce on the
star they orbit; these result in small, but detectable, periodic
wavelength shifts in the stellar spectrum.  Radial
velocity favors the detection of massive, Jupiter-class
objects orbiting close to the star, leaving open the
question of whether the architecture of our solar system in which
the giant planets occupy orbits $\geq$ 5 AU from the star is common
or rare.  This distinction is important since it is believed that
giant planets cannot form close to a star and must spiral in
from a much more distant radius, disrupting the stable
orbits of any terrestrial planets in their path.  In this
way, the arrangement of giant planets around a star is related to the
probability that that star harbors Earth-like planets in stable orbits
at radii that allow for the presence of liquid water and possibly
life.

\section{ESPI Scientific Objectives}

The scientific objective of the baseline 3-year ESPI mission is to
directly image and characterize extra-solar giant planets in $\geq 5$
AU-radius orbits around 160-175 candidate stars, all of which are
brighter than apparent magnitude V = 8 and lie within 16 parsecs of
Earth; five of the nearest stars also offer an opportunity for
detecting terrestrial-type planets.  As shown in  Figure 1,
ESPI will successfully image planets whose orbital periods range up to
30 years, providing unique access to the outer regions of exoplanetary
systems.


Most of these systems could be observed in 18 months, leaving 18
months for follow-up studies to confirm common proper motion (used to
distinguish a planet from a chance background source) and to fit
partial orbits to derive orbital parameters (in conjunction with
precise radial velocity measurements).  ESPI will also obtain
spectroscopic and photometric measurements of several of the brightest
planets found.  

A variety of different conditions and chemical processes in the
atmospheres of extra-solar planets is anticipated, making spectroscopy
in the 5000 to 10,000 $\AA$ region essential, even if only a
handful of the ESPI-discovered planets are sufficiently bright to
permit it.  If Jupiter-like planets with orbital radii $\geq$ 5 AU are
common, the ESPI mission may yield two or three dozen detections of
planets toward which filter spectrophotometry should be possible, and
a further dozen with sufficient signal-to-noise ratios, SNR, to permit
low resolution ($\lambda/\Delta\lambda \leq$ 40) spectroscopy at SNR
$\sim$ 8 to 9.  Even these limited capabilities will permit ESPI to
distinguish the albedo of planets with spectra similar to that of
Jupiter, as contrasted to Uranus.  The methane band at 9000 $\AA$
and the continuum breakbetween Jupiter-like gas giants and
Uranus-like ice giants near 6000 $\AA$ are clearly detectable.

\section{Technical Approach}

ESPI uses an Apodized Square Aperture (ASA) to achieve the high
dynamic range required for exoplanet imaging.  The ESPI optical design
employs an unobscured primary mirror in an off-axis Cassegrain
configuration. An ASA transmission mask is located close to the focal
plane along with a coronagraphic blocker at the focus. The ASA mask
has the form of the product of two crossed prolate spheroidal (PS)
functions. The PS functions are chosen to minimize diffraction close
to the central peak and to maximize overall transmission through the
mask(Papoulis, 1973). The ASA technique is described in detail in
Nisenson and Papaliolios, 2001, Ap. J., 548, L201.

Figure 2 shows (a) Cuts through two versions of prolate
spheroid apodization functions. (b) Three point spread functions from
PS apodization adjusted for different throughputs and diffraction
suppression: 48\% (top), 33\% (middle), and 23\% (bottom). (c) Effect of
random transmission errors on the psf. (d) Effect of random phase
errors (flat spectrum at mid-frequencies, $1/f^2$ at higher
frequencies).  In panels b, c, and d, off-diagonal separations of $\pm
3\lambda / D$ are shown with vertical dashed lines.


The ESPI dynamic range will be limited by scattered light from the
optical surfaces, not diffraction. Figure 3 is a result of an
accurate computer simulation showing how ESPI could detect planetary
systems: (a) The PS apodized aperture. (b) Two giant planets as seen
with ESPI (no wavefront aberration) from 10 pc with the inner planet
1/2 Jupiter diameter and 2.5 AU from star and the outer planet 1
Jupiter at 5 AU from star.  Note the suppression of diffraction
everywhere except along the central cross (which is blocked). (c) ESPI
image with 1/1000 wave optical quality for 5th magnitude star, 1 hour
integration. (d) Image after subtraction of a field with no
planets. For spectroscopy, rotation of the telescope allows locating a
dark speckle in the position of the planet, improving the signal-
to-noise ratio of the observation.

We have also tested the ESPI apodization approach in the laboratory.
Using pairs of point sources and a superpolished imaging mirror, we
have demonstrated detecting contrasts of $3 \times 10^9$ between the
``star'' and a ``planet'' separated by $6 \times \lambda /D$ is
possible.


\section{Summary}

ESPI has been proposed for direct imaging and
spectral analysis of giant planets orbiting solar-type stars. It also
permits unique observations of many Galactic, extragalactic and
cosmological sources.  ESPI has an off-axis cassegrain design with a
square telescope mirror. The apodization mask is located near the
telescope focus and is optimized for transmission and for the
narrowness of the central peak of the PSF, since this sets the angular
resolution of the system. ESPI would be capable of detecting
Jupiter-like planets in relatively long-period orbits around as many
as 160 to 175 stars with a signal-to-noise ratio $\geq$ 5.  In
addition to the survey, ESPI will also study a few of the brightest
discovered planets spectroscopically and spectro-photometrically to
distinguish ice giants like Uranus and Neptune from gas giants like
Jupiter and Saturn, and to determine whether super-Earth and
super-Venus planets exist.

\section{Figure Captions}

\noindent Figure 1 - Discovery Space for ESPI

\noindent Figure 2 - Apodizing Functions and PSF's for ESPI

\noindent Figure 3 - Simulation of ESPI Planet Detection

\end{document}